\documentclass[structabstract]{aa}  
\usepackage{graphicx}
\usepackage{txfonts}
\usepackage{natbib}
\begin{document}
\title{Filament Fragmentation in High-Mass Star Formation\thanks{Based on
    observations carried out with the IRAM Plateau de Bure
    Interferometer. IRAM is supported by INSU/CNRS (France), MPG
    (Germany) and IGN (Spain). The data are available in electronic
    form at the CDS via anonymous ftp to cdsarc.u-strasbg.fr
    (130.79.128.5) or via
    http://cdsweb.u-strasbg.fr/cgi-bin/qcat?J/A+A/}.}


   \author{H.~Beuther
          \inst{1}
          \and
          S.E.~Ragan
          \inst{2}
          \and
          K.~Johnston
          \inst{2}
          \and
          Th.~Henning
           \inst{1}
          \and
          A.~Hacar
           \inst{3}
           \and
          J.T.~Kainulainen
           \inst{1}
           }
   \institute{$^1$ Max Planck Institute for Astronomy, K\"onigstuhl 17,
              69117 Heidelberg, Germany, \email{name@mpia.de}\\
              $^2$ University of Leeds, Leeds, LS2 9JT, UK\\
              $^3$ University of Vienna, T\"urkenschanzstr. 17, A-1180 Vienna, Austria}

   \date{Version of \today}

\abstract
{Filamentary structures in the interstellar medium are crucial
  ingredients in the star formation process. They fragment to form
  individual star-forming cores, and at the same time they may also
  funnel gas toward the central gas cores providing an additional gas
  reservoir.}
{We want to resolve the length-scales for filament formation and fragmentation
  (resolution $\leq$0.1\,pc), in particular the Jeans length and
  cylinder fragmentation scale.}
{We have observed the prototypical high-mass star-forming filament
  IRDC\,18223 with the Plateau de Bure Interferometer (PdBI) in the
  3.2\,mm continuum and N$_2$H$^+$(1--0) line emission in a ten field
  mosaic at a spatial resolution of $\sim 4''$ ($\sim$14000\,AU).}
{The dust continuum emission resolves the filament into a chain of at
  least 12 relatively regularly spaced cores. The mean
      separation between cores is $\sim$0.40$(\pm 0.18)$\,pc. While
      this is approximately consistent with the fragmentation of an
      infinite, isothermal, gravitationally bound gas cylinder, a high
      mass-to-length ratio of $M/l\approx
      1000$\,M$_{\odot}$\,pc$^{-1}$ requires additional turbulent
      and/or magnetic support against radial collapse of the
      filament. The N$_2$H$^+(1-0)$ data reveal a velocity gradient
  perpendicular to the main filament. Although rotation of the
  filament cannot be excluded, the data are also consistent with the
  main filament being comprised of several velocity-coherent
  sub-filaments.  Furthermore, this velocity gradient
  perpendicular to the filament resembles recent results toward
  Serpens south that are interpreted as signatures of filament
  formation within magnetized and turbulent sheet-like
  structures. Lower-density gas tracers ([CI] and C$^{18}$O) reveal a
  similar red/blueshifted velocity structure on scales around $60''$
  east and west of the IRDC\,18223 filament. This may tentatively be
  interpreted as a signature of the large-scale cloud and the
  smaller-scale filament being kinematically coupled. We do not
  identify a velocity gradient along the axis of the filament. This
  may either be due to no significant gas flows along the filamentary
  axis, but it may partly also be caused by a low inclination angle of
  the filament with respect to the plane of the sky that could
  minimize such signature.}
{The IRDC\,18223 3.2\,mm continuum data are consistent with thermal
  fragmentation of a gravitationally bound and compressible gas
  cylinder. However, the large mass-to-length ratio requires
  additional support -- likely turbulence and/or magnetic fields --
  against collapse. The N$_2$H$^+$ spectral line data indicate a
  kinematic origin of the filament, but we cannot conclusively
  differentiate whether it has formed out of (pre-existing)
  velocity-coherent sub-filaments and/or whether magnetized converging
  gas flows, a larger-scale collapsing cloud or even rotation played a
  significant role during filament formation.}

\titlerunning{Filament Fragmentation in High-Mass Star Formation}

\maketitle

\section{Introduction}
\label{intro}

\begin{figure*}[htb] 
\begin{center}
\includegraphics[width=0.99\textwidth]{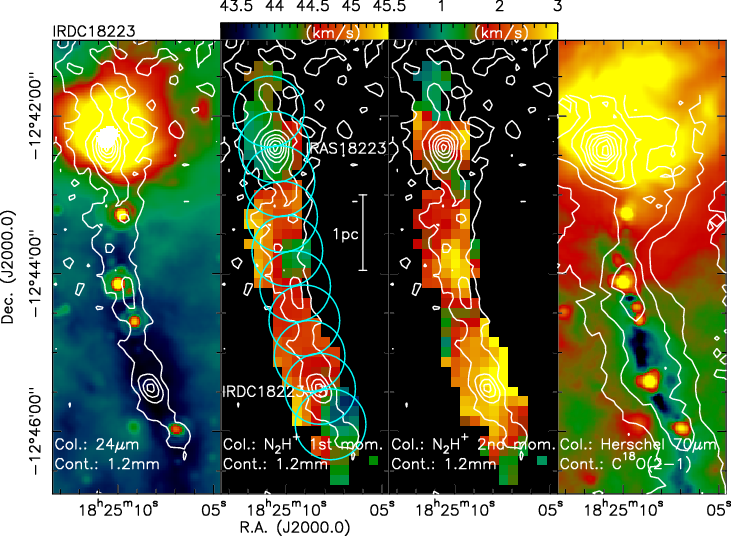}
\end{center}
\caption{The filamentary cloud IRDC\,18223: The left panel presents
  the Spitzer 24\,$\mu$m image in color \citep{beuther2007a}, and the
  MAMBO 1.2\,mm continuum map in contours \citep{beuther2002a} with
  levels from 5 to 95\% of the peak emission of 753\,mJy\,beam$^{-1}$
  (in 10\% steps).  The 2nd and 3rd panels present the Nobeyama 45\,m
  N$_2$H$^+$(1--0) 1st and 2nd moment maps, respectively
  \citep{tackenberg2014}.  The right panel shows the Herschel
  70\,$\mu$m observations (in color) and in contours the IRAM 30\,m
  C$^{18}$O(2--1) data integrated from 40 to 49\,km\,s$^{-1}$.  The
  contours are from 5 to 95\% of the peak emission of
  18.6\,K\,km\,s$^{-1}$ (in 10\% steps). Additionally, the middle
  panel shows the two main sources (the HMPO IRAS18223-1243 in the
  north and IRDC18223-3 in the south), a scale bar, as well as the
  positions of the 10 observed mosaic fields.}
\label{18223}
\end{figure*}

The existence of filaments in the interstellar medium has been known
for a long time. Especially the arrival of the Herschel observatory
has strongly increased the interest in filamentary structures, in
particular as filaments are a main evolutionary stage during the
formation of dense cores and stars
\citep{andre2010,henning2010,menshchikov2010,arzoumanian2011,andre2014}.
During star formation, filaments can fragment and form the seeds of
the star-forming cores.  Furthermore, gas can be funneled along the
filaments and feed the star-forming regions (e.g.,
\citealt{schneider2010,kirk2013,myers2013b,tackenberg2014}). This is
especially important for high-mass star formation because massive
stars are likely fed from the larger-scale environment (e.g.,
\citealt{smith2009b}).  Several filament studies have revealed
filament parameters like their density structure, stability criteria,
fragmentation length, characteristic width or kinematic properties
(e.g.,\,\citealt{johnstone2003b,jackson2010,schneider2010,henning2010,beuther2011b,hacar2013,hill2012,kainulainen2013,henshaw2014,kainulainen2015}).
While large-scale low-mass star-forming regions have recently studied
at high spatial resolution with CARMA (e.g.,
\citealt{fernandez2014,lee2014}), many previous studies in the
high-mass regime discussed spatial structures (on the order of
0.2\,pc) based largely on single-dish measurements. Prominent
exceptions are the infrared dark cloud investigated by
\citet{battersby2014}, the hierarchical accretion study by
\citet{galvan2010} or the recent ALMA studies by, e.g.,
\citet{peretto2013} or \citet{zhang2015}. The obvious goal is to
investigate such filamentary structures at small spatial scales to
study the physical properties of filaments on the scales of core and
star formation on the order of 10000\,AU. One recent high-resolution
study of a massive filament has been performed by \citet{henshaw2014}
who do not find large velocity gradients across the filament but
rather resolve it into velocity-coherent sub-structures, similar to
the results of \citet{hacar2013} for a low-mass filament in
Taurus\footnote{While \citet{hacar2013} required filaments to be
  connected within sonic velocity separation, \citet{henshaw2014} were
  less restrictive and allowed connected filaments based on the
  measured FWHM.}.

With the ultimate goal to understand high-mass star formation in
filaments forming out of the larger-scale cloud, here we study the
$\sim$4\,pc long infrared-dark filament associated with the High-Mass
Protostellar Object (HMPO) IRAS\,18223-1243
\citep{sridha,beuther2002a}.  This region is part of a very long
($>$50\,pc) filament previously investigated on larger scales by
\citet{kainulainen2011}, \citet{tackenberg2013}, \citet{ragan2014},
 and \citet{zucker2015}.  The region is a well studied high-mass
star formation complex at a distance of $\sim$3.5\,kpc encompassing
various evolutionary stages from very young protostars embedded in an
infrared dark cloud (IRDC) to forming HMPOs (e.g.,
\citealt{sridha,garay2004,beuther2002a,beuther2005d,beuther2007a,beuther2010b,fallscheer2009}).
Based on Spitzer, Herschel and mm single-dish data, the mass and
luminosity distribution of the filament is known in detail (e.g.,
\citealt{beuther2010b,ragan2012b}), and we now aim to shift the focus
to the smaller scale dust and gas kinematic properties to set these
into context with the larger-scale dust and gas filament.

For the region of this study, we have an extensive set of
complementary data comprising -- among others -- the full Herschel
far-infrared continuum data from 70 to 500\,$\mu$m and additional
ground based longer wavelength data \citep{beuther2010b,ragan2012b},
high-density N$_2$H$^+$(1--0) data from the Nobeyama 45\,m telescope
\citep{tackenberg2014}, as well as [CII], [CI] and CO observations
from SOFIA, APEX and the IRAM\,30m telescope
\citep{beuther2014}. Figure \ref{18223} presents a compilation of 
different continuum and spectral line data of this region. However,
while all these observations address the large-scale structure of the
gas on scales of $\sim$0.3\,pc, no data existed so far that can
investigate the fragmentation properties and kinematics of the region
on scales comparable to the Jeans length.  For example, average
densities of high-mass star-forming regions around $10^5$\,cm$^{-3}$
at low temperatures of 15\,K result in typical Jeans fragmentation
scales of around 10000\,AU.  These latter scales are now resolved with
our new Plateau de Bure Interferometer 3\,mm line and continuum data.

The main scientific questions to be addressed are: What is the
relevant fragmentation scale of the filament itself? Is it consistent
with Jeans fragmentation, isothermal gaseous cylinders or even more
complex structures? What are the kinematic properties of the gas? Does
it show velocity gradients along the filament indicative of global
collapse (e.g., \citealt{tackenberg2014}) or rather velocity coherent
sub-filamentary structures (e.g., \citealt{hacar2013,henshaw2014})?
What are the virial parameters of the gas cores? What is the
mass-per-unit length of the filament?

\section{Observations} 
\label{obs}

The IRDC\,18223 filament was observed during a series of 8 
tracks between June 2013 and April 2014 in the D- (with 5 antennas)
and C-array (with 6 antennas) configurations of the Plateau de Bure
Interferometer (PdBI).  The projected baselines ranged between 15 and
175\,m. While the absolute reference position was R.A.~(J2000.0)
18h25m09.533s and Dec.~(J2000.0) $-$12$^o$43$'$55.90$''$, ten mosaic
pointings were required to cover the length of the filament
(Fig.~\ref{18223}). The adopted velocity of rest $v_{\rm{lsr}}$ of the
system is 45.3\,km\,s$^{-1}$. Bandpass calibration was conducted with
either of 3C279, 2200+420, 1633+382 or 3C345. The absolute flux
calibration was performed with MWC349 and is estimated to be correct
to within $\sim$15\%.  Phase and amplitude calibration was conducted
with regular observations of the quasars 1730-130 and 1741-038. The
spectral coverage of the wide-band receiver and correlator unit ranged
from 91.53 to 95.14\,GHz. Almost the whole bandpass was used to
extract the 3.2\,mm continuum emission (a small band around the
N$_2$H$^+(1-0)$ was excluded). The narrow-band correlator units
focused mainly on the N$_2$H$^+(1-0)$ and $^{13}$CS(2--1) lines. The
nominal channel separation was 0.039\,MHz, and we smoothed the data to
0.2\,km\,s$^{-1}$ spectral resolution for our final data-cubes. While
the N$_2$H$^+(1-0)$ emission was easily detected throughout the whole
filament and will be discussed in detail in this paper,
$^{13}$CS(2--1) was only detected toward the strong northern IRAS
source and will not be discussed any further. During the imaging
process, we experimented with different weighting schemes between
natural and uniform weighting to optimize the spatial resolution as
well as signal-to-noise ratio. For the continuum emission we present
both data with resulting synthesized beams of $5.6''\times 3.27''$ (PA
$-6^o$) and $4.37''\times 2.84''$ (PA $+187^o$), respectively.  For
the N$_2$H$^+(1-0)$ data, we discuss the results based on the
naturally weighted data with a synthesized beam of $5.77''\times
3.39''$ (PA $-6^o$). The $1\sigma$ rms values are
0.13\,mJy\,beam$^{-1}$ for the continuum and 9\,mJy\,beam$^{-1}$
measured in a line-free channel of the N$_2$H$^+(1-0)$ emission.

Although Fig.~\ref{18223} presents single-dish N$_2$H$^+(1-0)$ data
from the Nobeyama 45\,m telescope, neither the spectral resolution nor
the sensitivity of these data is sufficient for merging with the new
interferometer data. Therefore, we refrain from doing that.

\section{Results}

The IRDC\,18223 represents a beautiful example of an infrared dark
filament within a larger-scale filamentary structure covering more
than 50\,pc in linear extent
\citep{kainulainen2011,tackenberg2013,ragan2014}. In the following we
will analyze in detail the dense gas and dust properties of this
filament from a fragmentation and a kinematic point of view.

\begin{figure}[htb] 
\includegraphics[width=0.49\textwidth]{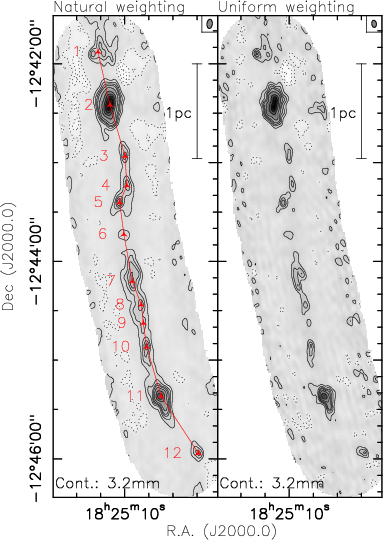}
\caption{3.2\,mm continuum images of the IRDC\,18223 filament. The
  left panel is imaged with natural weighting and a resulting
  synthesized beam of $5.6''\times 3.27''$\, with a position angle of
  $-6^o$ whereas the right image is conduced with uniform weight and a
  synthesized beam of $4.37''\times 2.84''$\, with a position angle of
  187$^o$. The contour levels in both images start at the $3\sigma$
  level of 0.39\,mJy\,beam$^{-1}$, continuing in $3\sigma$ steps until
  1.56\,mJy\,beam$^{-1}$ (the same are shown as dotted lines for negative
  features). Then they continue from 2.34\,mJy\,beam$^{-1}$ in
  1.56\,mJy\,beam$^{-1}$ steps. The red markers and line in the left panel
  outline the main continuum peak positions and guide the eye for the
  filament structure. The synthesized beams and scale bars are shown in
  both panels at the top-right.}
\label{continuum}
\end{figure}

\subsection{The filament in the 3.2\,mm continuum emission}
\label{cont}

Figure \ref{continuum} presents the 3.2\,mm continuum data imaged with
natural and uniform weighting. While the uniform weighting reveals
more small-scale detail, the naturally-weighted imaged recovers more
of the extended filamentary structure. Considering our spatial
resolution limit of $4.37''\times 2.84''$ that corresponds at the
given distance of 3.5\,kpc to a linear resolution element of
$\sim$\,13000\,AU, we cover scales of the filament between several pc
down to $\sim$0.063\,pc. With only a small large-scale velocity
gradient along the filament (Sec.  \ref{kinematics}), IRDC\,18223 may
either have barely any gas motions along the filament or have only a
small inclination angle with respect to the plane of the sky. With an
extent of $\geq$4\,pc and an approximate width of on average less than
0.2\,pc, its length-over-width ratio exceeds $\sim$20.

\begin{figure*}[htb] 
\begin{center}
\includegraphics[width=0.99\textwidth]{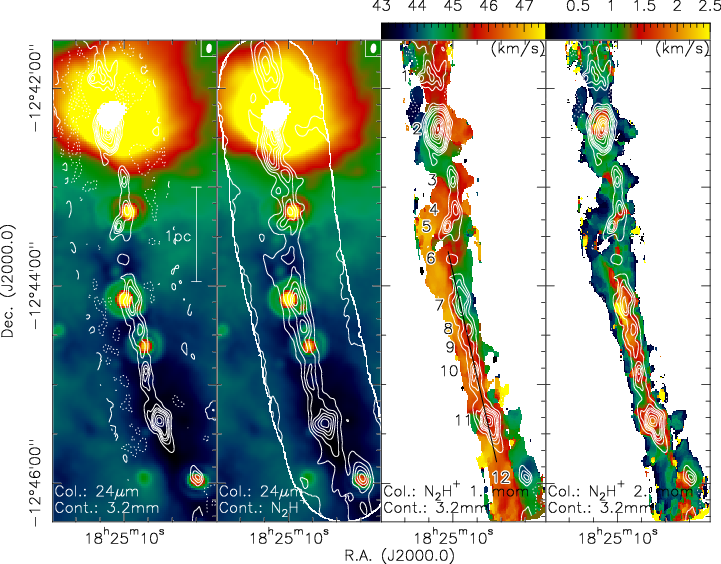}
\end{center}
\caption{The left and 2nd panel show the Spitzer 24\,$\mu$m emission in
  color with a stretch to increase the visibility of the mid-infrared
  dark filament. The left panel presents in contours the 3.2\,mm
  continuum emission with natural weighting and the same levels as in
  Fig.~\ref{continuum}. For comparison, the 2nd panel shows in
  contours the integrated N$_2$H$^+(1-0)$ emission covering all
  hyperfine structure components between 35 and 55\,km\,s$^{-1}$. The
  third and 4th panel present the 1st and 2nd moment maps
  (intensity-weighted peak velocities and line-width) extracted from
  the isolated most negative hyperfine structure component. In the 3rd
  panel, the velocities are shifted by the separation of the hyperfine
  structure line of 8\,km\,s$^{-1}$ to the $v_{\rm{lsr}}$. The
  contours in the 3rd and 4th panel both show the same continuum
  contours as in the left panel. The black line in panel 3 gives the
  direction of the position-velocity cut shown in Fig.~\ref{pv}. A
  scale-bar is depicted in the left panel.}
\label{n2h+}
\end{figure*}

Fits to the spectral energy distributions of Spitzer and Herschel data
toward individual cores in that region resulted in temperatures
usually between 15 and 20\,K for the cold component in this filament
\citep{beuther2007a,beuther2010b,ragan2012b}. For the following
estimates we assume 15\,K throughout the whole studied region.
Assuming optically thin dust continuum emission at this temperature
with a dust opacity $\kappa_{1.07{\rm mm}}\sim 0.8$\,cm$^2$\,g$^{-1}$
at densities of $10^5$\,cm$^{-3}$ \citep{ossenkopf1994} and a
gas-to-dust mass ratio of 150 \citep{draine2011}, the MAMBO 1.2\,mm
single-dish continuum data from \citet{beuther2002a} give an estimate
of $\sim$4000\,M$_{\odot}$ total mass within the filament outlined in
Figure \ref{18223}\footnote{Typical errors are dominated by the dust
  model and temperature uncertainties and range within a factor
  $\sim$2 (e.g., \citealt{ossenkopf1994,beuther2002a}.}. Since
bolometer arrays like MAMBO also filter large-scale emission, we can
use the Herschel 500\,$\mu$m data from
\citet{beuther2010b}\footnote{For this new estimate we used a newly
  reduced data product provided by the EPOS (Early Phase Of Star
  Formation) Key-Project team.} to estimate the upper mass limit for
the region. Integrating the Herschel 500\,$\mu$m flux over the whole
area shown in Fig.~\ref{continuum}, we get a total flux of
187\,Jy. With the same assumptions as above ($\kappa_{500\mu{\rm
    m}}\sim 3.9$\,cm$^2$\,g$^{-1}$ at densities of $10^5$\,cm$^{-3}$,
\citealt{ossenkopf1994}), this results in a total gas mass in this
area of $\sim$7673\,M$_{\odot}$, close to twice as much as measured by
the MAMBO 1.2\,mm data. Hence, the derived masses between 4000 and
7673\,M$_{\odot}$ bracket the overall mass of this star-forming
large-scale filament.  For comparison, the total 3.2\,mm flux of the
filament shown in the naturally-weighted image of Fig.~\ref{continuum}
is $\sim$149\,mJy which corresponds with the same assumptions
($\kappa_{3.2{\rm mm}}\sim 0.17$\,cm$^2$\,g$^{-1}$) to a total mass of
$\sim$2200\,M$_{\odot}$.  Hence, in this case even naturally-weighted
interferometer data recover between 29\% to 55\% of the single-dish
flux and hence resemble the overall structure very well.

\begin{table*}[htb]
  \caption{Fluxes, masses, column densities (at 15\,K) and luminosities of the individual cores}
\begin{tabular}{lrrrrrrrr}
  \hline \hline
  \# & R.A. & Dec. & $S_{\rm{peak}^a}$ & $S_{\rm{int}^a}$ & $N_{\rm{H_2}}$ & $M_{\rm{peak}}^e$ & $M$ & $L^c$\\
  & (J2000.0) & (J2000.0) & $\left(\frac{\rm{mJy}}{\rm{beam}}\right)$ & (mJy) & ($\frac{10^{23}}{\rm{cm^2}}$) & (M$_{\odot}$) & (M$_{\odot}$) & (L$_{\odot}$) \\
  \hline
  1$^b$ & 18:25:11.11 & -12:41:52.8 & 1.5  & 8.3  & 2.3  & 22 & 123 & 157 \\
  2 & 18:25:10.61 & -12:42:25.0 & 10.3 & 57.0 & 16.0 & 150 & 843 & 1979 \\
  2$^d$ &             &             &      &      & 7.2  & 77 & 377 &      \\
  3 & 18:25:09.99 & -12:42:56.1 & 1.4  & 3.1  & 2.2  & 20 & 46  \\
  4 & 18:25:09.91 & -12:43:14.0 & 1.5  & 4.1  & 2.3  & 22 & 61 & 135  \\
  5 & 18:25:10.21 & -12:43:24.2 & 2.0  & 5.1  & 3.1  & 29 & 75  \\
  6 & 18:25:10.04 & -12:43:43.6 & 0.6  & 0.9  & 0.9  & 9 & 13  \\
  7 & 18:25:09.69 & -12:44:12.1 & 1.8  & 12.0 & 2.8  & 26 & 177 & 201 \\
  8 & 18:25:09.31 & -12:44:26.4 & 1.6  & 6.3  & 2.5  & 23 & 93 \\
  9 & 18:25:09.23 & -12:44:37.6 & 1.0  & 2.4  & 1.6  & 15 & 36 \\
  10 & 18:25:09.09 & -12:44:52.0 & 1.7  & 6.9  & 2.6  & 25 & 102  \\
  11 & 18:25:08.49 & -12:45:22.0 & 6.7  & 26.0 & 10.4 & 97 & 385 & 324 \\
  12 & 18:25:06.94 & -12:45:56.7 & 1.3  & 2.4  & 2.0  & 19 & 36 & 79 \\
  \hline \hline
\end{tabular}
{\footnotesize ~\\
  $^a$ Peak and integrated fluxes are extracted from the naturally-weighted 3.2\,mm continuum  image within the $3\sigma$ contours.\\
  $^b$ This is the approximate mid-point between the three sub-peaks there.\\
  $^c$ From \citet{ragan2012b}.\\
  $^d$ Calculated also at 31\,K following \citet{beuther2010b}.\\
  $^e$ Peak masses $M_{\rm{peak}}$ are calculated from the peak column densities $N_{\rm{H_2}}$ for a more accurate comparison with virial masses measured toward the peak positions (Table \ref{fits}).}
\label{masses}
\end{table*}

\begin{figure*}[htb] 
\begin{center}
\includegraphics[width=0.99\textwidth]{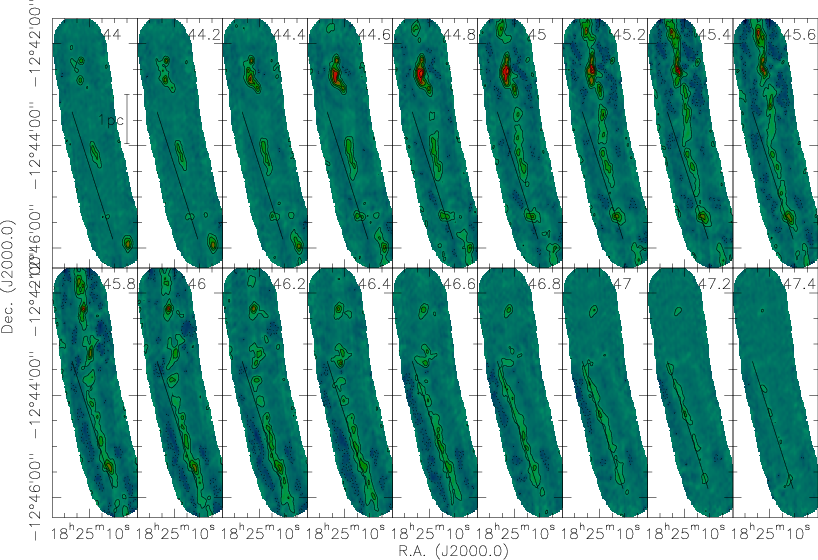}
\end{center}
\caption{Channel map of the isolated component of the N$_2$H$^+(1-0)$
  line. The velocities are shifted by the separation of the hyperfine
  structure line of 8\,km\,s$^{-1}$ to the $v_{\rm{lsr}}$. Velocities
  for each channel are given in the top-right of each panel. The
  contours in each panel are from 5 to 95\% of the respective peak
  emission in that channel. Negative features at the same contour
    levels are shown in dashed lines. A linear scale bar is presented
    in the top-left panel. The line is meant to guide the eye and is
  drawn along the main filamentary structure visible in the channel at
  47.2\,km\,s$^{-1}$.}
\label{channel}
\end{figure*}

We extracted also the peak $S_{\rm{peak}}$ and integrated fluxes
$S_{\rm{int}}$ of the 12 substructures identified in
Fig.~\ref{continuum} (left panel), which are listed in Table
\ref{masses}. These 12 regions were identified by eye with the goal to
encompass major structures along the filament. For source
identification and flux extraction we used mainly the naturally
weighted image (Fig.~\ref{continuum} left panel). We restricted
ourselves to structures that are detected above a $4\sigma$ level also
in the uniformly-weighted image in Fig.~\ref{continuum} (right panel).
Obviously, some of these regions can fragment at smaller scales which
is already visible for the northernmost peak 1, or also for peak 7 in
the uniformly-weighted image. Also extended structures in
Fig.~\ref{continuum} (left panel), e.g., between peaks 6 and 7 may be
separate fragments (see Fig.~\ref{continuum} right panel), however,
they cannot be resolved as individual peaks in the naturally weighted
image and are hence not separately extracted.  Emission was integrated
above the 3$\sigma$ level of the naturally-weighted image
(Fig.~\ref{continuum} left panel), the separation between sources were
estimated by eye, usually in emission troughs approximately in the
middle between adjacent cores. While the peak positions and hence peak
column densities are well defined (Table \ref{masses}), defining exact
boundaries between cores is less straightforward (e.g.,
\citealt{kainulainen2009b,pineda2009}). Nevertheless, in this rather
linear structure of the IRDC\,18223 filament, small shifts of
boundaries between cores only marginally affect the integrated fluxes
and hence estimated masses.  Assuming optically thin dust continuum
emission at 15\,K temperature, the masses $M$ and column densities
$N_{\rm{H_2}}$ corresponding to $S_{\rm{int}}$ and $S_{\rm{peak}}$ are
also compiled in Table \ref{masses}.  Furthermore, for a better
comparison with the virial masses below we calculate the peak masses
$M_{\rm{peak}}$ from the peak column densities $N_{\rm{H_2}}$ over the
size of the beam (Table \ref{fits}). For region 2, we also calculate
the masses and column densities at a higher temperature of 31\,K as
derived in \citet{beuther2010b}.  The fragment masses span a broad
range between 13 and 843\,M$_{\odot}$, and the column densities range
between $9\times 10^{23}$ and $1.6\times 10^{24}$\,cm$^{-2}$,
corresponding to visual extinctions between approximately 100 and
1000\,mag, well within the regime of the precursors of high-mass stars
(e.g., \citealt{kauffmann2010}).  Table \ref{masses} also gives the
luminosities derived for individual cores based on Herschel data by
\citet{ragan2012b}. These luminosities range from starless cores
without any far-infrared detections and hence no measurable
luminosities, via low-luminosity sources around 100\,$L_{\odot}$ up to
almost 2000\,L$_{\odot}$ from the high-mass protostellar object source
2. One should keep in mind that the measured luminosities are
bolometric luminosities and do not necessarily stem from hydrogen
burning but are likely still dominated by accretion luminosity. The
corresponding mass and column density sensitivities estimated from our
$3\sigma$ flux level of 0.39\,mJy\,beam$^{-1}$ are
$\sim$5.8\,M$_{\odot}$ and $\sim$6.1$\times 10^{22}$\,cm$^{-2}$,
respectively. Two of the regions are well-studied star-forming
objects: core 2 corresponds to IRAS\,18223-1243 \citep{sridha} and
core 10 is the younger high mass protostellar object IRDC18223-3
\citep{beuther2007a,fallscheer2009}.

Assuming spherical symmetry within our beam size, the above estimated
$3\sigma$ column density sensitivity of $\geq$6.1$\times
10^{22}$\,cm$^{-2}$ results in an approximate density sensitivity of
$\geq$2.6$\times 10^5$\,cm$^{-3}$. Combining this density sensitivity
with the missing flux ratio, we can estimate a dense gas mass fraction
above that density threshold between $\sim$29\% and 55\%. Assuming for
the dense gas a star formation efficiency of $\sim$30\% (e.g.,
\citealt{alves2007,andre2014}), this results in an approximate overall
star formation efficiency of the whole filamentary cloud of 8.7\% to
16.5\%. These latter values depend strongly on the star formation
efficiency which may vary between 20\% and 40\% \citep{andre2014}.

\begin{table}[htb]
\caption{Peak separations between mm continuum peaks}
\begin{tabular}{lr}
\hline \hline
peak pairs & peak separations\\ 
& pc \\
\hline
1-2 & 0.56 \\
2-3 & 0.55 \\
3-4 & 0.30 \\
4-5 & 0.19 \\
5--6 & 0.33 \\
6--7 & 0.49 \\
7--8 & 0.26 \\

8--9 & 0.19 \\
9--10 & 0.24 \\
10--11 & 0.53 \\
11--12 & 0.70\\ 
\hline \hline
\end{tabular}
\label{sep}
\end{table}

Another important parameter for the fragmentation processes is the
separation between the fragments along the filament. While individual
gas cores along the filament may fragment on even smaller scales
beyond our spatial resolution and mass sensitivity limits, here we
focus on the fragmentation of the filament itself. Table \ref{sep}
shows the projected separations between the 12 cores identified in
Fig.~\ref{continuum} and Table \ref{masses}. These values result in a
mean projected separation between the fragments of $\sim$0.40\,pc with
a standard deviation of 0.18\,pc. This mean separation should be
considered as an upper limit because more fragments along the filament
may exist that were not identified within the spatial resolution and
mass sensitivity of our data. For comparison, we can estimate the
Jeans length for mean densities of the large-scale filament between
$10^4$ and $10^5$\,cm$^{-3}$ at 15\,K to a range between $\sim$0.23
and $\sim$0.07\,pc. These Jeans scales are considerably smaller than
the observed  mean fragment separation. Although the
  measured fragment separations are projected upper limits, a
  difference of more than a factor of 2 between measurements and Jeans
  lengths appears significant. We will discuss this result in the
context of filament fragmentation in section \ref{fragmentation}.

Furthermore, the mass-to-length ratio $M/l$ can be compared to the
critical mass per length $M/l_{\rm{crit}}$ in an equilibrium situation
(e.g., \citealt{ostriker1964,fiege2000a,fiege2000b}).  Deviations from
this $M/l_{\rm{crit}}$ indicate non-equilibrium modes. While filaments
with line masses below $M/l_{\rm{crit}}$ expand if not supported by
external pressure, filaments with line masses strongly exceeding
$M/l_{\rm{crit}}$ can collapse radially perpendicular to the main long
axis of the filament. Such radial collapse would be on very short
time-scales and hence prohibit further fragmentation along the main
axis of the filament (e.g., \citealt{inutsuka1992}). If turbulent
pressure dominates over thermal pressure, using
$\sim$2.5\,km\,s$^{-1}$ as an approximate line width from the
single-dish data for the dense gas of the IRDC\,18223 filament
(Fig.~\ref{18223}, third panel 2nd moment map), one can estimate a
critical mass to length ratio $(M/l)_{\rm{crit-turb}}\approx 84\times
(\delta v)^2\approx 525$\,M$_{\odot}$\,pc$^{-1}$ (e.g.,
\citealt{jackson2010}), more than 20 times higher than the
$(M/l)_{\rm{crit-therm}}\approx 25$\,M$_{\odot}$\,pc$^{-1}$ for an
undisturbed filament using the thermal sound speed at 15\,K of
0.23\,km\,s$^{-1}$ \citep{ostriker1964}.

  For comparison, we can estimate the $M/l$ of the IRDC\,18223
  filament by dividing the total filament mass estimated from the
    1.2\,mm MAMBO data of $\sim$4000\,M$_{\odot}$ by the projected
  length of the total filament of $\sim$4\,pc. This results in a very
  high $M/l$ of 1000\,M$_{\odot}$\,pc$^{-1}$. Although larger than the
  estimated $(M/l)_{\rm{crit-turb}}$, the difference of about a factor
  2 is within the uncertainties considering that the linewidth is
  squared in the equation above. Hence, turbulent motions may help
  stabilizing the filament against fast radial collapse. One should
  keep in mind that turbulence can also create shocks and by that
  density enhancements and filaments. In addition to this, magnetic
  fields may help to stabilize the filament as well. For example,
  recent magneto-hydrodynamic (MHD) simulations by \citet{kirk2015}
  showed that in their MHD case the $M/l_{\rm{crit-B}}$ increased by a
  factor of $\sim$3 compared to the pure hydro case. Although
    radial collapse cannot be entirely excluded, the fact of the
    existence of this filament with such a regular fragment separation
    as well as the large $(M/l)_{\rm{crit-turb}}$ make radial collapse
    a less likely scenario.

\subsection{The gas kinematics of the filament}
\label{kinematics}

Figure \ref{n2h+} presents a compilation of the N$_2$H$^+(1-0)$ emission in
comparison to the 3.2\,mm cold dust continuum as well as the
24\,$\mu$m mid-infrared warm dust emission/absorption. The dense gas
tracer N$_2$H$^+$ follows the filamentary structure very well
with the integrated emission peaks showing close resemblance to the
peak positions in the 3.2\,mm continuum (Fig.~\ref{n2h+}, 1st and 2nd
panel). The line emission seems even to trace the larger-scale
structure slightly better than the continuum observations.

\begin{figure*}[htb] 
\includegraphics[width=0.24\textwidth]{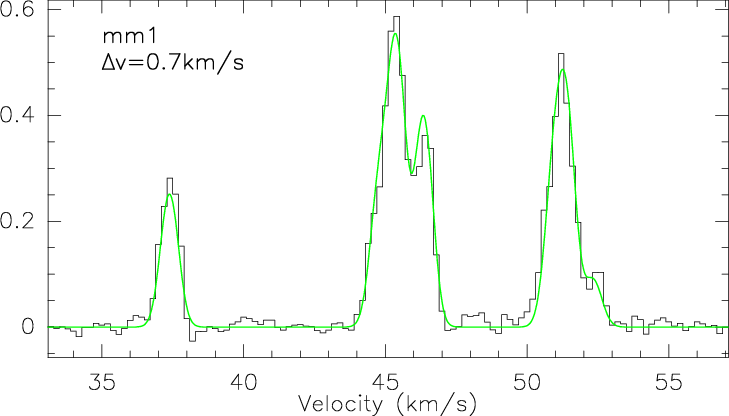}
\includegraphics[width=0.24\textwidth]{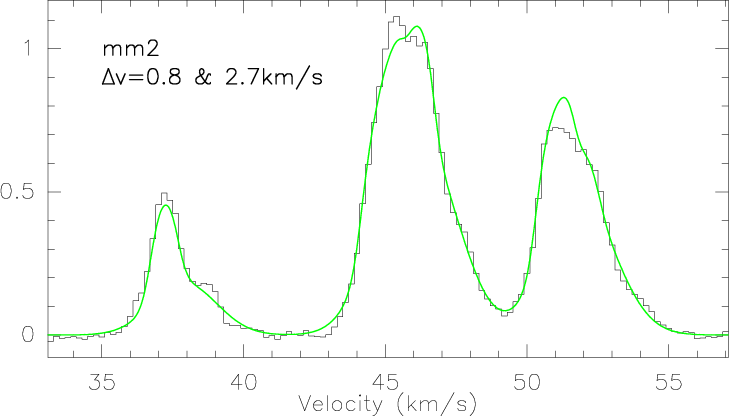}
\includegraphics[width=0.24\textwidth]{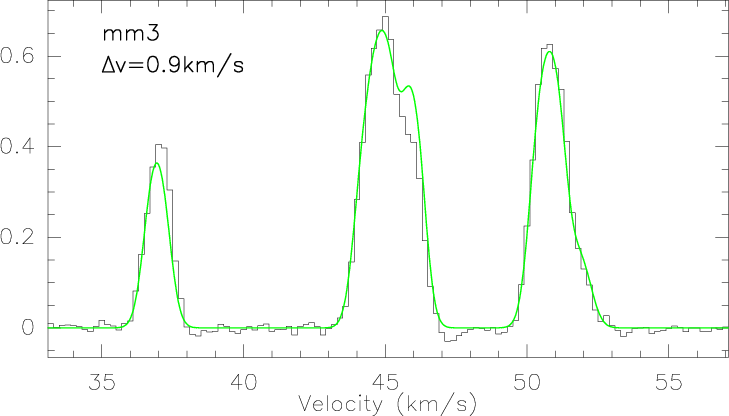}
\includegraphics[width=0.24\textwidth]{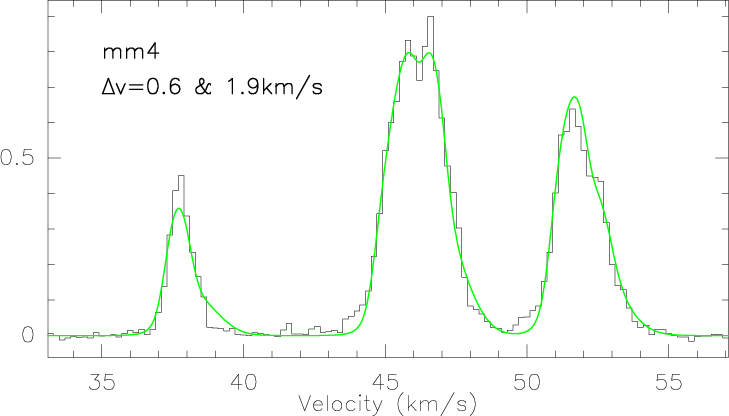}\\
\includegraphics[width=0.24\textwidth]{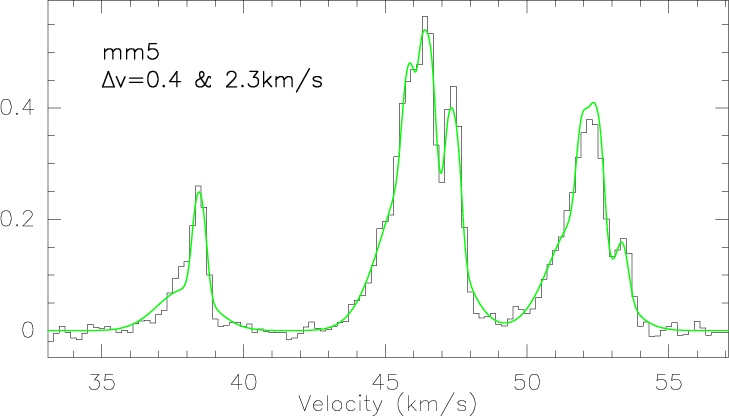}
\includegraphics[width=0.24\textwidth]{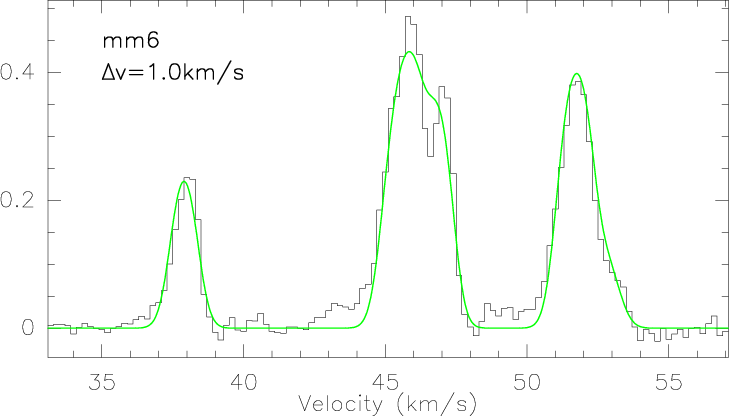}
\includegraphics[width=0.24\textwidth]{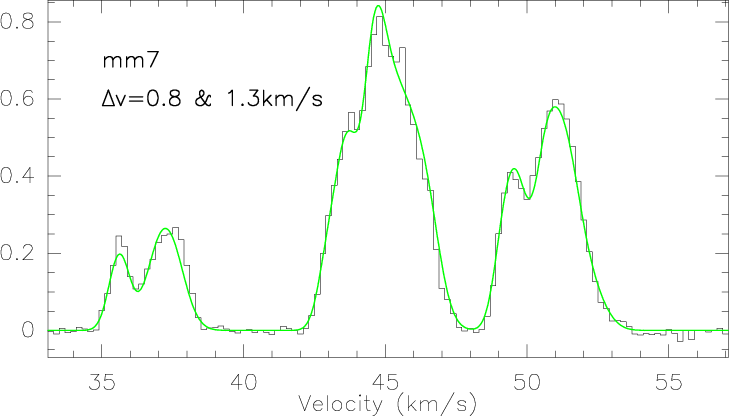}
\includegraphics[width=0.24\textwidth]{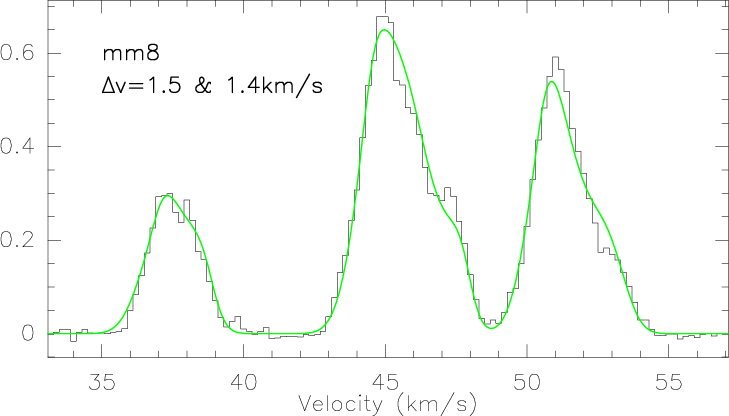}\\
\includegraphics[width=0.24\textwidth]{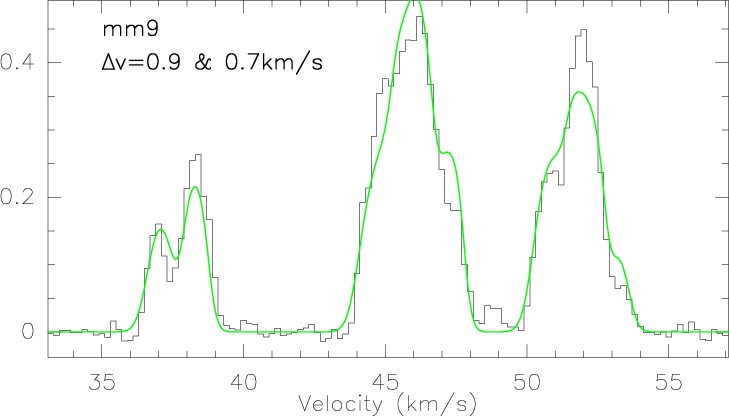}
\includegraphics[width=0.24\textwidth]{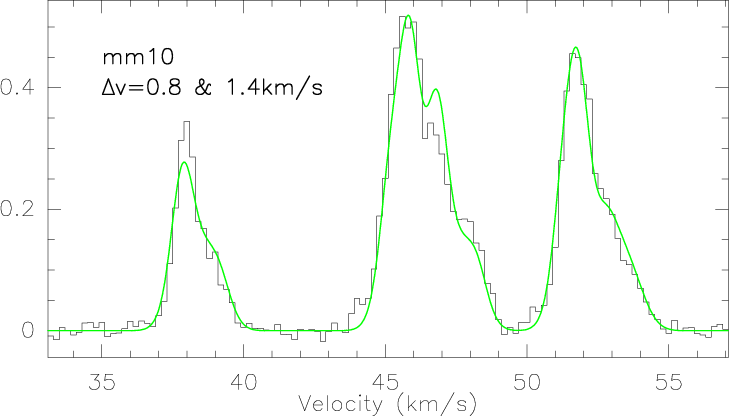}
\includegraphics[width=0.24\textwidth]{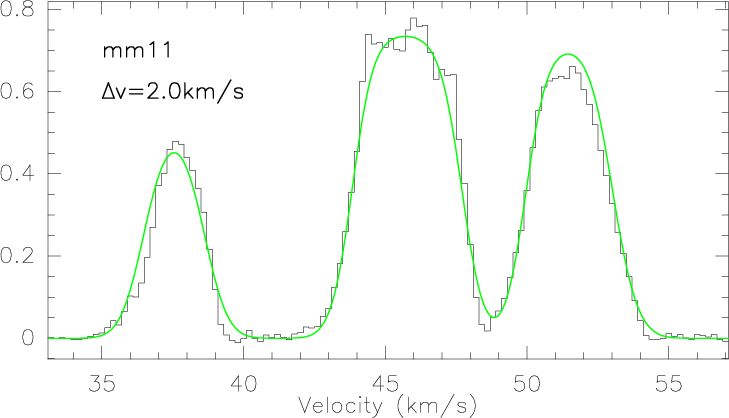}
\includegraphics[width=0.24\textwidth]{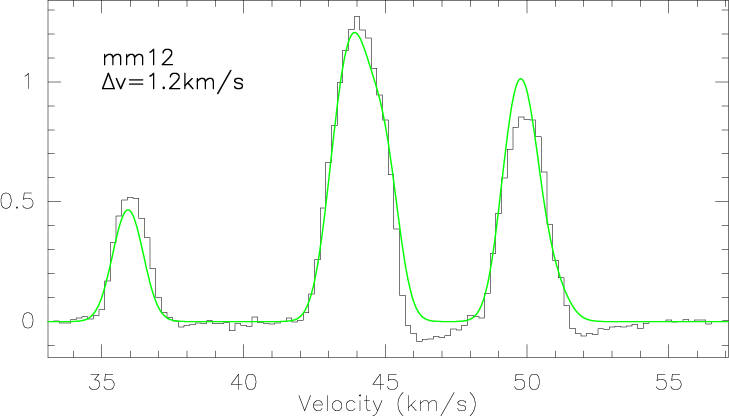}
\caption{Spectra and hyperfine structure line fits towards the 12
  positions marked in Fig.~\ref{continuum} and Table \ref{masses}. The
  y-axis is in units of Jy\,beam$^{-1}$. The full width half maximum
  values of single- or double-component fits are shown in each panel.}
\label{spec}
\end{figure*}

The third panel in Fig.~\ref{n2h+} presents a first moment map
(intensity-weighted peak velocities) of the whole filament (conducted
with the isolated hyperfine structure component shifted by
8\,km\,s$^{-1}$ to the $v_{\rm{lsr}}$). While the combined optical
depth of all hyperfine components together often exceeds 1 toward the
main core peak positions (e.g., Fig.~\ref{spec}), the moment maps are
done with the isolated hyperfine component 8\,km\,s$^{-1}$ apart from
the main component at a relative intensity of $\sim$11\%, hence at
optical depth $\ll 1$. Therefore, the peak velocities in
Fig.~\ref{n2h+} should not be affected by optical depth effects. The
northern part of the region between peak 1 and 5 does shows some
variation between red- and blue-shifted gas without any clear
signature along or across the filament. In particular, one can
identify a clear velocity gradient across core 2 in the
north. However, since that is a well-known high-mass protostellar
object likely driving a molecular outflow \citep{sridha}, this
velocity structure is likely strongly influenced by the dynamics of
the already more evolved internal source.  The situation is different
from peak 6 southward. In this southern part, the first moment map is
indicative of a velocity gradient across the filament from red-shifted
gas in the east to blue-shifted gas in the west.  Conducting a cut
perpendicular to the filament through continuum source 8 in
Fig.~\ref{n2h+}, we measure peak velocities between 44.8 and
47.0\,km\,s$^{-1}$ over an angular size of $\sim$15$''$, corresponding
to $\sim$0.26\,pc at the given distance, corresponding to a velocity
gradient of $\sim$25.6\,km\,s$^{-1}$\,pc$^{-1}$. The velocity gradient
can also be seen by inspecting the channel map of the gas
(Fig.~\ref{channel}). The almost vertical line in the channel map is
drawn to guide the eye along the filamentary structure seen in the
channel at 47.2\,km\,s$^{-1}$. This channel map also shows that at
redshifted velocities the gas in the south is located further to the
east than at blue-shifted velocities.  An exception is the emission of
peak 10 (also known as IRDC\,18223-3) which covers almost the entire
velocity range. This peak is clearly more evolved than the other
fragments and also drives an energetic outflow \citep{fallscheer2009}.
Fig.~\ref{n2h+} also presents a second moment map (intensity-weighted
line width) which shows a line width increase toward the ridge of the
filament often closely related to the mm continuum emission. While
some of this increased line width is clearly associated with star
formation activity (e.g., peaks 2 and 10 also known as IRAS18223-1243
and IRDC\,18223-3), in other cases, this second moment increase can
also be caused by multiple velocity components (see below).

One can now ask whether this more than 2\,pc long southern part of the
filament is either a velocity-coherent structure with only a small
velocity-gradient from east to west, or whether it may consist of
several sub-filaments that may form the large structure (e.g.,
\citealt{hacar2013,smith2014}). To analyze this in more detail,
Fig.~\ref{spec} presents the N$_2$H$^+(1-0)$ spectra extracted toward
the 12 mm peak positions. For all spectra we fitted the full hyperfine
structure with either one or two spectral velocity components. The
decision to fit either single or multiple components was mainly based
on the shape of the isolated most-blueshifted hyperfine structure line
component. The fit results are shown in Table \ref{fits} and in Figure
\ref{spec}. More than 50\% of the spectra clearly need two velocity
components to adequately fit the data (e.g., mm2 and mm7, even the
spectrum toward peak 10 is likely an overlap of multiple components
potentially caused by the internal outflow-driving source).  It is
interesting to note that most of the single component fits as well as
often one of the two components exhibit relatively narrow line width
below 1\,km\,s$^{-1}$.  Nevertheless, although comparably narrow, even
these components have significant non-thermal contributions since the
thermal line width of N$_2$H$^+$ at 15\,K is $\sim$0.15\,km\,s$^{-1}$.
The second component usually is slightly broader between 1.3 and
2.7\,km\,s$^{-1}$.  These broader components may also be comprised out
of multiple components along the line of sight.

The peak velocities extracted from the spectra do not exhibit a clear
velocity gradient along the filament. Therefore, in contrast to the
velocity-gradient seen in the moment map from east to west, there is
no obvious velocity gradient from north to south along the
filament. As already mentioned for the whole filament, this missing
gradient along the filament may either be due to missing significant
streaming motions along, or it may also be introduced due to a
possible orientation almost in the plane of the sky.

The measured line width toward the individual peak positions can also
be used to get a rough estimate of the virial mass of the individual
fragments. Following \citet{maclaren1988}, we estimate the virial mass
according to $M_{\rm{vir}} = k_2 \times R \times \Delta v^2$, where we
use $k_2=126$ for a $\rho\propto r^{-2}$ density distribution,
$R=16000$\,AU$\sim 0.078$\,pc the size of the region corresponding to
the beam size of the N$_2$H$^+(1-0)$ data, and the measured full width
half maximum $\Delta v$. The resulting virial masses $M_{\rm{vir}}$
are shown in Table \ref{fits}. A direct correlation with the masses
derived from the dust emission in Table \ref{masses} is not possible
because some sources have more than one N$_2$H$^+$ component and thus
several virial mass estimates, whereas the gas masses from the
continuum emission are always single values. Nevertheless, in most
cases of multiple velocity components, one dominates. Table \ref{fits}
also shows the ratio of the virial masses divided by the peak masses
$M_{\rm{peak}}$ derived from the peak column densities $N_{\rm{H}_2}$
in Table \ref{masses}. While in a majority of cases this
$M_{\rm{vir}}/M_{\rm{peak}}$ ratio is considerably smaller than 1,
there are also several cases where the ratio is close to unity or even
higher. Interestingly, the three highest $M_{\rm{vir}}/M_{\rm{peak}}$
sources 4, 5, and 6 show very different internal luminosity
characteristics. While source 4 has a clear associated internal
heating source, \#5 is at the edge of that, and \#6 exhibits no
detectable 24\,$\mu$m emission (Fig.~\ref{n2h+}). From a virial
balance point of view, this comparison is an indicator that many of
the individual cores are prone to collapse or already collapsing (see
also recent filament work by \citealt{battersby2014}). The latter is
also evident by embedded sources within many of the dust continuum
peaks (e.g., Fig.~\ref{18223} left and right, mid- to far-infrared
panels). However, there are exceptions which may still be stable
against gravitational collapse (e.g., source 6).

\begin{table}[htb]
\caption{N$_2$H$^+(1-0)$ hyperfine structure line fits}
\begin{tabular}{lrrrr}
\hline \hline
\# & $v_{\rm{peak}}$ & $\Delta v$ & $M_{\rm{vir}}$ & $\frac{M_{\rm{vir}}}{M_{\rm{peak}}}^b$ \\
   & (km\,s$^{-1}$)  & (km\,s$^{-1}$) & (M$_{\odot}$)\\
\hline
1 & 45.4 & 0.7 & 4.8 & 0.22\\
2 & 45.2 & 0.8 & 6.3 & $-^c$ \\
  & 45.9 & 2.7 & 71.3 & 0.92 \\
3 & 44.9 & 0.9 & 7.9 & 0.36 \\
4 & 45.7 & 0.6 & 3.5 & $-^c$ \\
  & 46.1 & 1.9 & 35.3 & 1.60 \\
5 & 45.9 & 2.3 & 51.7 & 1.78 \\
  & 46.4 & 0.4 & 1.6 & $-^c$\\
6 & 45.9 & 1.0 & 9.8 & 1.09 \\
7 & 43.6 & 0.8 & 6.3 & 0.24 \\
  & 45.2 & 1.3 & 16.5 & 0.63 \\
8 & 44.8 & 1.5 & 22.0 & 0.96 \\
  & 45.9 & 1.4 & 19.2 & 0.82 \\
9 & 45.1 & 0.9 & 8.0 & 0.53 \\
  & 46.3 & 0.7 & 6.3 & 0.42 \\
10 & 45.9 & 0.8 & 6.3 & 0.25 \\
  & 46.4 & 1.4 & 19.2 & 0.77 \\
11& 45.6 & 2.0 & 39.1 & 0.40 \\
12& 43.9 & 1.2 & 14.1 & 0.74 \\
\hline \hline
\end{tabular}
\label{fits}
{\footnotesize ~\\
$^a$Peak velocities and FWHM values from fits to full hyperfine structure\\
$^b$ $M_{\rm{peak}}$ are taken from the continuum data in Table \ref{masses}.\\
$^c$ The line widths and virial masses of these secondary velocity components are negligible compared to the main components.}
\end{table}

\begin{figure}[htb] 
\includegraphics[width=0.49\textwidth]{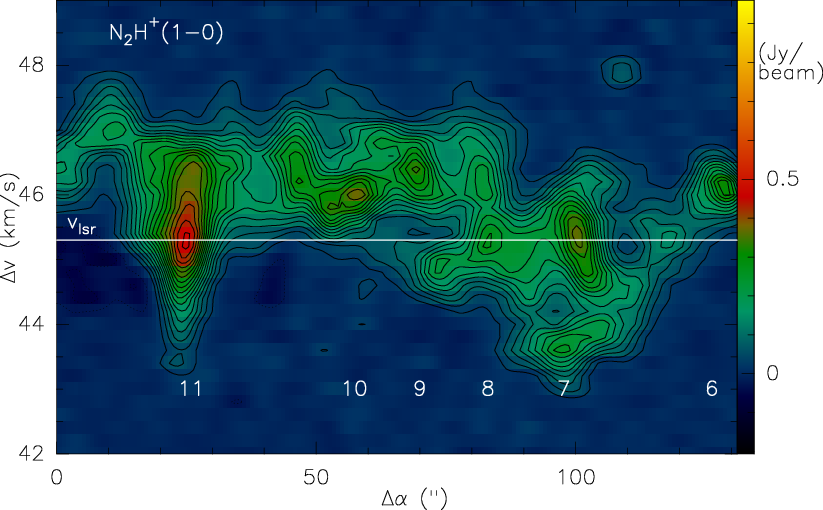}
\caption{Position-velocity cut through the southern half of the
  filament in the isolated N$_2$H$^+(1-0)$ line (shifted to the
  $v_{\rm{lsr}}$) from south to north along the axis shown as black
  line in panel 3 of Fig.~\ref{n2h+}. The mm peak names are marked
  at the bottom.}
\label{pv}
\end{figure}

A different way to look at the velocity structure is a
position-velocity cut along the southern filament. Figure \ref{pv}
presents such a position-velocity (pv) diagram going from south to
north between peaks 11 and 6 as outlined in Figure \ref{n2h+}. While
this pv cut exhibits very broad emission in the south toward peak 11
(or IRDC\,18223-3) which is also visible in Figs.~\ref{channel} \&
\ref{spec}, the rest of the cut does show a variety of features: while
some peaks exhibit single emission peaks redshifted with respect to
the $v_{\rm{lsr}}$ (e.g., peak 10 or 6), others show multiple peaks on
both sides of the $v_{\rm{lsr}}$ (e.g., peaks 7 to 9). Again, no clear
velocity gradient across the structure can be identified. However, the
position-velocity cut shows that internal structure is found within
this filament. As outlined in section \ref{cont}, our identified
fragments within the filament (Table \ref{masses}) represent likely
only a lower limit to the actual structures, and more fragments may
exist. For example, in Fig.~\ref{pv} we can identify structures
between peaks 6 and 7.  Comparing these
to the continuum maps in Fig.~\ref{continuum}, they are spatially
close to the extended structures in Fig.~\ref{continuum} (left panel)
and the $4\sigma$ peaks in Fig.~\ref{continuum} (right panel).

\begin{figure}[htb] 
\includegraphics[width=0.49\textwidth]{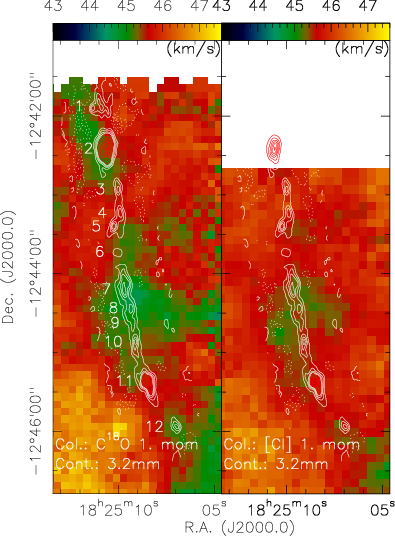}
\caption{The two panels show for comparison the first moment maps in
  molecular C$^{18}$O(2--1) and atomic carbon [CI] obtained from the
  IRAM\,30m and APEX data presented in \citet{beuther2014}. The
  contours are the 3.2\,mm continuum emission with natural weighting
  and the same levels as in Fig.~\ref{continuum}. The numbers of the
  mm cores are given in the left panel.}
\label{other}
\end{figure}

\begin{figure*}[htb] 
\includegraphics[width=0.99\textwidth]{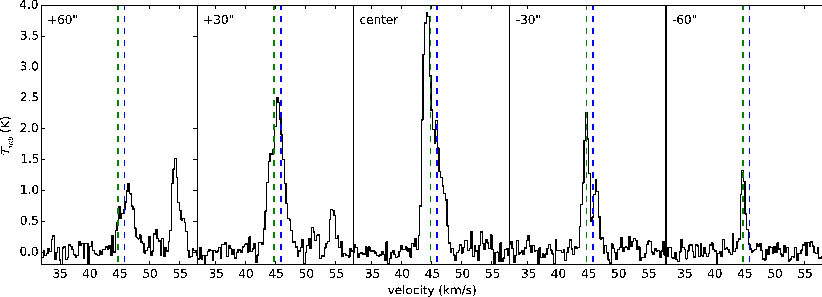}
\caption{C$^{18}$O(2--1) spectra from the IRAM\,30m telescope
  extracted along a cut perpendicular to the filament at a declination
  of mm continuum source 8 (\citealt{beuther2014}, Ragan et al.~in
  prep.). The green and blue vertical lines mark the two velocity
  components measured in N$_2$H$^+$(1--0) toward the central source 8.
  The R.A.~offsets are marked in each panel.}
\label{spec_c18o}
\end{figure*}

A natural question is whether similar kinematic features can be found
in the less dense surrounding of the gas on larger spatial scales. In
a recent [CII]/[CI]/C$^{18}$O emission line study toward four IRDCs,
\citet{beuther2014} identified several velocity components between 42
and 56\,km\,s$^{-1}$ in comparably localized emission structures in
the atomic and molecular carbon emission east of the IRDC\,18223
filament. To investigate this in more detail, we used the atomic [CI]
and molecular C$^{18}$O(2--1) data of the region
\citep{beuther2014}. Figure \ref{other} shows the first moment maps of
these tracers toward our target region.  In these moment maps, for
both tracers no obvious large-scale velocity gradients across the
filament can be identified. The only mentionable feature in
Fig.~\ref{other} is that the dense filament seem in C$^{18}$O and [CI]
appears to be slightly blue-shifted compared to the larger scale
molecular cloud. This effect is more pronounced in the C$^{18}$O(2--1)
than the [CI] emission.  For comparison, we extracted the
C$^{18}$O(2--1) spectra toward five positions across the filament at
the declination of mm continuum source 8
(Fig.~\ref{spec_c18o}). Toward the central positions, these spectra
show that the two N$_2$H$^+$ velocity components are approximately
recovered in the C$^{18}$O emission as well. This indicates that, if
the region were observed at higher angular resolution also in
C$^{18}$O, one would likely identify similar velocity structures as in
N$_2$H$^+$. Moving outward to the east, the emission becomes more
redshifted whereas to the west the emission is more blueshifted. This
is particularly prominent in the spectra at $+60''$ and $-60''$
(corresponding to $\sim\pm$1\,pc, Fig.~\ref{spec_c18o}), respectively.
Interestingly, this velocity shift found in the individual spectra is
hard to identify in the moment maps because several components overlap
diminishing the signatures this way. It should be noted that the
velocity gradient found this way on large scales for the cloud is
considerably smaller than the 25.6\,km\,s$^{-1}$\,pc$^{-1}$ found
above for the dense filament in the interferometric N$_2$H$^+$
data. While it appears in general plausible that the velocities
increase when going to higher densities, recent molecular cloud and
filament formation simulations tend to find similar signatures (e.g.,
\citealt{moeckel2015,smith2015}). This trend needs to be investigated
more in the future from an observational as well as theoretical point
of view.  Nevertheless, the general trend of redshifted gas in the
east and blueshifted gas in the west is found on the large scales
traced by the C$^{18}$O single-dish emission as well as on the small
scales studied with the new PdBI N$_2$H$^+$ data. Hence, we find
tentative evidence that the large-scale cloud and the smaller-scale
filament are kinematically coupled.

\section{Discussion}

\subsection{Filament fragmentation}
\label{fragmentation}

As outlined in Section \ref{cont}, the separations between the main
identified fragments along the filament exceed the Jeans length at the
given densities and temperatures. Although our measured projected
separations are upper limits because of potential unresolved
fragmentation and/or fragments below our sensitivity limit, the
difference between Jeans length and projected fragment separation by
more than a factor two appears significant.  As a next step, we
analyze the region in the framework of isothermal, gravitationally
bound gaseous cylinders. Based on early work of
\citet{chandrasekhar1953,nagasawa1987} and \citet{inutsuka1992} as
well as more recent adaptions like \citet{jackson2010},
\citet{beuther2011b} or \citet{kainulainen2013}, we study the
conditions of an infinite isothermal gas cylinder. Although
IRDC\,18223 is obviously not of infinite length, the filament is part
of a much larger structure extending more than 50\,pc projected on the
sky. Hence, in this context, the approximation of IRDC\,18223 being
part of a much longer, almost infinite structure seems justifiable.
Furthermore, since star formation has already started at different
locations in the filament, it is not the perfect starless filament
anymore. However, the infrared dark nature of IRDC\,18223 clearly
shows the youth of the whole structure. Hence, it is still an
excellent target region that represents conditions in the relatively
early phase of filament fragmentation.

This characteristic isothermal scale-height $H$ of such a gas cylinder
is given by $H = c_s(4\pi G\rho)^{-1/2}$ with $c_s$ the sound speed,
$G$ the gravitational constant and $\rho$ the gas mass density at the
center of the filament (e.g., \citealt{nagasawa1987}). With a thermal
sound speed of the gas at 15K of $c_s \sim 0.23$\,km\,s$^{−1}$ and
using an approximate density $\rho \sim 10^5$\,cm$^{-3}$ (e.g.,
\citealt{beuther2002a}), the characteristic scale-height is $H\sim
0.02$\,pc. In that case, the characteristic fragmentation scale
corresponding to the fastest growing unstable mode of the fluid
instability is $\lambda_{\rm{frag}} = 22H$ (e.g.,
\citealt{nagasawa1987,inutsuka1992,jackson2010}) which results with
the above used parameters for the sound speed and the temperature in
an approximate fragmentation scale $\lambda_{\rm{frag}} \approx
0.44$\,pc. One should keep in mind that there are uncertainties
associated with the assumed density and sound speed, which introduce
an uncertainty in the estimated core separation.  In particular,
instead of the thermal sound speed of the gas, other works have used
the Gaussian turbulent velocity dispersion in this estimate, which
would increase the estimated value for $\lambda_{\rm{frag}}$ according
to the ratio of the turbulent velocity dispersion over the thermal
sound speed (e.g., \citealt{wang2014}).  Nevertheless, it is
interesting to note that using the thermal sound speed at 15\,K and an
approximate central density of $10^5$\,cm$^{-3}$ results in
$\lambda_{\rm{frag}}$ which is close to the mean core separation of
the filament of $\sim$0.40\,pc discussed in section
\ref{cont}. Although our observed mean separation is likely only an
upper limit (Section \ref{cont}), an isothermal, gravitationally bound
and compressible gas cylinder allows us to reproduce the general
fragmentation structure of IRDC\,18223 well.

The high mass-to-length ratio estimated from the dense gas
  bolometer single-dish data of $M/l$ of
$\sim$1000\,M$_{\odot}$\,pc$^{-1}$ exceeds by far the critical mass to
length ratios for thermal filaments, but it is approximately
consistent with a scenario where also turbulent support of the
filament against radial collapse is considered (Section \ref{cont}).
Additional effects, e.g., the presence of magnetic fields may add
further support to the stability of the filament. The recent MHD
simulations by \citet{kirk2015} revealed an increase of the critical
mass-to-length ratio by a factor of $\sim$3 compared to their pure
hydrodynamical simulations. One should keep in mind that only if the
filament does not collapse radially perpendicular to the filament,
perturbations along the filament can grow and result in fragmentation
as discussed above (e.g., \citealt{inutsuka1992}). Quantitatively
speaking, it is one of the largest $M/l$ values found so far, where
Nessie is reported with $\sim$110\,M$_{\odot}$\,pc$^{-1}$, Orion with
385\,M$_{\odot}$\,pc$^{-1}$ \citep{bally1987,jackson2010}, a
sub-filament in the G35.39 region with 115\,M$_{\odot}$\,pc$^{-1}$
\citep{henshaw2014}, and the G11.11 IRDC with a value of
600\,M$_{\odot}$\,pc$^{-1}$ \citep{kainulainen2013}.

\subsection{Kinematics of the filament}

The kinematic properties of filaments have recently been discussed in
a different context. Signatures of gas flows along the filaments have
been identified (e.g.,
\citealt{hacar2011,kirk2013,tackenberg2014,zhang2015}), and velocity
coherent sub-structures have been identified that may originate from
the filament formation process (e.g.,
\citealt{hacar2013,smith2014,henshaw2014}). Furthermore, recently
velocity gradients perpendicular to filaments were discussed that may
also be signatures of filament formation \citep{fernandez2014}.

Potential signatures for gas flows along the filament are
blue-red-shifted velocity structures on the two opposite sites of mm
peak positions, indicating that the gravitational well of the gas
peaks attracts the gas from the filament. While we cannot exclude such
a possibility, the moment maps and position-velocity diagrams
(Figs.~\ref{n2h+} \& \ref{pv}) at least show no obvious signatures of
this. The fact that we do not see any strong gradient from
north to south could be due either to barely any streaming motions
along the filament, or it can also indicate that the filament may be
oriented relatively close to the plane of the sky. Such a low
inclination angle would make any identification of gas flows along the
filament very difficult.

Another signature of gas flows along the filament could be regular
oscillatory-like velocity changes along such a filament as discussed
in \citet{hacar2011} or \citet{zhang2015}. While the pv-cut in Figure
\ref{pv} shows velocity fluctuations around the $v_{\rm{lsr}}$, it is
hard to identify such a sea-saw pattern. Therefore, this signature is
also not evident in our data.

In a recent study of the filaments in Serpens south,
\citet{fernandez2014} find velocity gradients perpendicular to
filaments as well, very similar to our findings here. Although they
also find velocity-gradients along the filaments, they stress that the
velocity-gradients perpendicular to the filaments are about an order
of magnitude larger than those along the filaments. The quantitative
gradients they derive perpendicular and parallel to the filament are
11.9\,km\,s$^{-1}$\,pc$^{-1}$ versus 0.9\,km\,s$^{-1}$\,pc$^{-1}$,
respectively.  This is suggestive for the kinematics perpendicular to
the filaments being dynamically much more important than those along
the filaments. The gradient we find in IRDC\,18223 is even larger with
$\sim$25.6\,km\,s$^{-1}$\,pc$^{-1}$. Although \citet{fernandez2014}
postpone a more detailed comparison with simulations to a future paper
(Mundy et al.~in prep.), the main thrust of their interpretation is
that such velocity-gradients perpendicular to filaments may stem from
motions associated with the formation and growing of the filaments
(see also \citealt{heitsch2013a,heitsch2013b}). Hence they may be
direct signatures of the filament formation processes. Similar results
are also indicated by recent three-dimensional simulations of
turbulence compressed regions in strongly-magnetized sheet-like layers
\citep{chen2015}. Within these layers, dense filaments and embedded
self-gravitating cores form via gathering or compression of the
material along the magnetic field lines. As a result of the mass
collection along preferred directions, velocity gradients
perpendicular to the filament major axis are a common feature seen in
their simulations \citep{chen2015}.

Regarding velocity-coherent sub-filaments, the velocity shifts of the
southern part of the filament, particularly prominent in the channel
map (Fig.~\ref{channel}), allows us to speculate whether we have also
in this region similar structures. While in principle rotation of the
filament is a possible explanation for the observed spectral
signatures, \citet{kirk2013} showed that for an accreting
  filament, infall motions would quickly dominate the kinematics, even
  if the filament were initially rotating. In Fig.~\ref{pv}, one can
tentatively identify several contiguous regions in
space/velocity between offsets $\sim 40''$ and $\sim 80''$ (peaks 9
and 8) and a velocity of $\sim 46$\,km\,s$^{-1}$, as well as between
$\sim 70''$ and $\sim 120''$ (midway between peaks 8/9 and midway
between peaks 6/7) and a velocity of $\sim 45$\,km\,s$^{-1}$.  While
\citet{hacar2013} find a large number of bundles of filaments
  within the Taurus region, \citet{henshaw2014} identify three
sub-filaments within their studied G35.39 dark filament. This latter
high-mass infrared dark cloud compares more closely to our IRDC\,18223
filament in physical properties (e.g., mass, size, velocity
dispersion) . In principle, a series of velocity-coherent
sub-filaments could mimic also a velocity gradient across the
structure. However, since the observations presented here are
interferometer-only data without the missing short-spacing
information, an automized search for velocity-coherent fibers in a
manner analogous to the studies by \citet{hacar2013} or
\citet{henshaw2014} is not possible. As mentioned before, the rms and
spectral resolution of our currently available Nobeyama 45\,m
single-dish data is not sufficient for a reasonable merging with the
interferometer data. However, we plan to obtain high-quality
short-spacing data in the future and revisit the analysis.

Alternative interpretations have recently been invoked to explain
observed trends in filament velocity structure. Recently,
\citet{tafalla2015} discussed the possibility that the filament in
Taurus which they observed may have formed from the supersonic
collision of sub-filaments, and that in the next step, this filament
will fragment due to gravitational instabilities. Similarly,
\citet{smith2014} recently discussed filament formation by
pre-existing structures based on hydrodynamic collapse simulations.

\begin{figure}[htb] 
\includegraphics[width=0.49\textwidth]{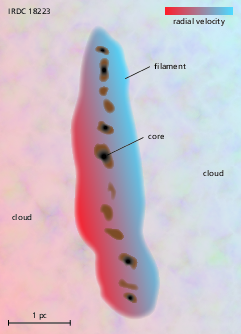}
\caption{Sketch of the approximate structure of IRDC\,18223.}
\label{sketch}
\end{figure}

An additionally interesting aspect in the presented results is that
the velocity gradient found on small scales in the interferometer data
perpendicular to the filament is also found on larger scales by the
single-dish C$^{18}$O(2--1) data (Fig.~\ref{spec_c18o}). This
indicates that larger-scale diffuse gas is kinematically coupled to
the dense inner filament. The fact that we find a significant steeper
velocity gradient in the dense inner filament compared to the
larger-scale cloud can also be identified in recent cloud and filament
formation simulations (e.g., \citealt{moeckel2015,smith2015}).

To summarize, on the spatial scales traced by our interferometer data, the
identified velocity-gradient perpendicular to the filament as well as
tentatively velocity-coherent sub-structures may both stem from
(magnetized) turbulent flows in sheet-like or filamentary
structures. Such turbulent and sheet-like substructures could have
formed out of the larger-scale self-gravitating and collapsing
cloud. Figure \ref{sketch} sketches the approximate structure of the
region highlighting the main features.

\section{Conclusions}
\label{conclusion}

The PdBI 3.2 mm line and continuum emission resolves this more than
4\,pc long filament into its substructures at about 15000\,AU scale.
We identify a linear structure with $\sim$12 cores at approximately
similar spacing with a mean projected separation of
$\sim$0.40($\pm$0.18)\,pc.  This separation is much larger than the
typical Jeans length. Although the observed core separation is an
upper limit because of limited spatial resolution and sensitivity, the
data are approximately consistent with the fragmentation properties of
an isothermal, gravitationally bound and compressible gas
cylinder. However, the mass-to-length ratio is also very high implying
additional turbulent and/or magnetic support of the filament if it is
supported against radial collapse.

We do not find any significant velocity gradient along the 4\,pc
filamentary structure, but the PdBI N$_2$H$^+(1-0)$ data reveal a
transverse velocity gradient across the southern half of the filament.
While it is possible that this southern filament may be composed of at
least two velocity-coherent sub-filaments, rotation of the filament
cannot be excluded either, although the latter appears less likely.
The missing signatures of gas flows along the filament may indicate
small streaming motions along the filament but they could also be
caused by a low inclination angle of the filament with respect to the
plane of the sky.

The velocity gradient perpendicular to the filament may also stem from
the filament formation process within magnetized and turbulent
sheet-like structures. On scales of $\pm 60''$ ($\sim\pm$1\,pc) east
and west of the filament we find similar red/blueshifted signatures as
on the smaller filament scales. This may be tentative evidence that
the lower-density cloud and higher-density filament indeed may be
kinematically coupled.

In summary, these combined line and continuum data reveal an excellent
example of a massive gas filament. While the continuum data are
roughly consistent with thermal fragmentation of a cylinder, the high
mass-to-length ratio requires additional support against radial
collapse, most likely due to turbulence and/or magnetic
fields. Furthermore, the observed velocity structure of the gas
indicates a dynamic origin of the filament. However, we cannot resolve
yet whether it is comprised of individual velocity-coherent
sub-filaments or whether other processes like magnetized converging
gas flows, a larger-scale collapsing cloud or even cloud rotation play
a significant role in the formation process. Future observations
combining single-dish and interferometer data will help to better
constrain these different scenarios.

\begin{acknowledgements} 

  We like to thank Axel Quetz for helping to produce the sketch in
  Fig.~\ref{sketch}. Furthermore, thanks a lot to Hendrik Linz
    from the EPOS Key-Project team for providing us a newly reduced
    version of the Herschel 500\,$\mu$m data. S.E.R. acknowledges
    support from VIALACTEA, a Collaborative Project under Framework
    Programme 7 of the European Union, funded under Contract \#
    607380.

\end{acknowledgements}

\bibliographystyle{aa}    


\end{document}